\documentclass[useAMS,usenatbib]{mn2e}
\usepackage{graphicx}

\title[AstraLux binary TEP hosts]{Stellar companions to exoplanet host stars: \\Lucky Imaging of transiting planet hosts\thanks{
Based on observations collected at the 2.2\,m telescope at Calar Alto Observatory, and on observations made with the 3.5\,m ESO NTT at La Silla 
Observatory under programme ID's 082.C-0084 and 083.C-0145.}}
\author[C. Bergfors et al. ]{C. Bergfors$^{1}$\thanks{E-mail:
bergfors@mpia.de}, W. Brandner$^{1}$, S. Daemgen$^{2}$, B. Biller$^{1}$, S. Hippler$^{1}$, M. Janson$^{3}$, \newauthor
N. Kudryavtseva$^{1}$, K. Gei{\ss}ler$^{4}$, T. Henning$^{1}$ and R. K\"ohler$^{1,5}$ \\
$^{1}$Max-Planck-Institut f\"ur Astronomie, K\"onigstuhl 17, 69117 Heidelberg, Germany\\
$^{2}$European Southern Observatory, Karl-Schwarzschild-Strasse 2, 85748 Garching, Germany\\
$^{3}$Princeton University, 4 Ivy Lane, Peyton Hall, Princeton, NJ 08544, USA\\
$^{4}$Dept. of Physics and Astronomy, State University of New York, Stony Brook, New York, 11794-3800, USA\\
$^{5}$Landessternwarte, Zentrum f\"ur Astronomie der Universit\"at Heidelberg, K\"onigstuhl, 69117 Heidelberg, Germany}
\begin{document}

\date{Accepted 2012 September 16. Received 2012 September 10; in original form 2012 May 21}

\pagerange{\pageref{firstpage}--\pageref{lastpage}} \pubyear{2012}

\maketitle

\label{firstpage}

\begin{abstract}
Observed properties of stars and planets in binary/multiple star systems provide clues to planet formation and evolution.
We extended our survey for visual stellar companions to the hosts of transiting exoplanets by 21 stars, using the Lucky Imaging technique with the two 
AstraLux instruments: 
AstraLux Norte at the Calar Alto 2.2-m telescope, and AstraLux Sur at the ESO 3.5-m New Technology Telescope at La Silla.
Typically a sensitivity to companions of magnitude 
difference $\Delta z^\prime\approx4$ is achieved at angular separation $\rho=0.5\arcsec$ and $\Delta z^\prime\ga6$ for $\rho=1\arcsec$.

We present observations of two previously unknown binary candidate companions, to the transiting planet host stars HAT-P-8 and WASP-12,
and derive photometric and astrometric properties of the companion candidates. 
The common proper motions of the previously discovered companion candidates with the exoplanet host stars TrES-4 and WASP-2 are confirmed from follow-up
observations. A Bayesian statistical analysis of 31 transiting exoplanet host stars observed with AstraLux suggests that the companion star
fraction of planet hosts is not significantly different from that of solar-type field stars, but that the binary separation is on average larger
for planet host stars.
\end{abstract}


\section{Introduction}
About half of solar-type stars in our neighbourhood are part of a binary or multiple system \citep{DuquennoyMayor1991, Raghavan2010}.
Understanding how a secondary star affects the formation and evolution of planets in the system is therefore of high importance
for an estimate of the overall occurrence of planets in our Galaxy. 

A close stellar companion is expected to 
affect planet formation in several ways, e.g., by heating and truncating the circumstellar protoplanetary disc \citep*{ArtymowiczLubow1994, Armitage1999, 
Nelson2000}, or by increasing the relative velocities of the planetesimals \citep{Heppenheimer1974, Heppenheimer1978, Whitmire1998}.
A secondary star may either stimulate \citep{Boss2006} or hinder \citep{Nelson2000, KleyNelson2008} planet formation.
The occurrence and properties of planets formed in binary systems may also provide a way to discriminate between the two most widely 
supported planet formation models: core accretion \citep[e.g., ][]{Pollack1996} and gravitational instability \citep[e.g., ][]{Boss1997,
Mayer2002}. 
For instance, formation by gravitational instability (GI) requires a massive disc for fragmentation to occur ($\rm M\ga0.1M_{\sun}$). 
\citet{Mayer2005} found that planet formation by GI in massive discs is not significantly different from formation around a single star if the binary
separation is $>120$ AU, while at close binary separations ($<60$ AU) the high temperatures caused by shock heating effectively suppress fragmentation.
Formation by core accretion, on the other hand, does not require a massive disc, and planets may form closer to their parent star without being much
affected by a close binary companion. Therefore, a trend of giant planet frequency with binary star separation, i.e., fewer giant planets in binary 
systems closer than 100 AU than in wider binaries or single systems, might point to GI as a main formation mechanism \citep{Mayer2005}.

Observations show that the frequency by which giant planets are formed in close binary stellar systems appears to be slightly
lower than around single stars. \citet{Eggenberger2008,Eggenberger2011} found that giant planets are more common in single-star systems than in
binaries separated by 35-100\,AU.
Nevertheless, surveys of multiplicity among planet host stars 
show that many exoplanet host stars are part of a binary or multiple system. We now know of $\sim50$
binary or multiple systems where the planet(s) belong to one of the stars (S-type orbit), and
Kepler observations recently revealed the first known planets that transit both stars in a close binary system, i.e., in a P-type orbit
\citep{Doyle2011, Welsh2012}.
System characteristics such as binary separation together with properties of the planets (orbital period, mass, eccentricity, etc.) and any differences
compared to the properties of the single-star planetary systems provide important constraints on 
planet formation and dynamical system evolution. 
Among suggested correlations between stellar and planetary properties in binaries, two appear significant: The most massive planets in 
short-period orbits belong to stars in binary systems, and so do the planets with the highest eccentricities 
\citep{DesideraBarbieri2007,SozzettiDesidera2010, Tamuz2008}.  

A significant fraction of exoplanets to date have been discovered using radial velocity measurements, and such systems have also been the subject of several multiplicity 
surveys \citep[e.g., ][]{Patience2002, Chauvin2006, Eggenberger2007, Mugrauer2007, Roberts2011, Mason2011, Ginski2012}.
However,
the discovery rate of transiting exoplanets (TEPs) has increased tremendously over the last couple of years, mainly due to the success of ground-based transit 
searches such as SuperWASP \citep{Pollacco2006} and HATNet \citep{Bakos2004}, as well as space-based programs such as CoRoT \citep{Baglin2006} and
Kepler \citep{Borucki2010}. Transiting exoplanets are unique in the way that properties such as mass and radius can be measured, and a variety of 
additional physical parameters such as true mass, mean density and surface gravity can thus be derived, potentially providing information on planet 
formation \citep*[see, e.g., ][]{Mazeh2005,Torres2008, Southworth2009, Southworth2010}. 
An unresolved faint secondary star within the photometric aperture, whether it is a bound companion or chance alignment, contributes
a constant flux offset to the transit light curve and affects the accuracy with which stellar and planetary parameters can be derived. High resolution
imaging of TEP hosts have shown that the presence of a (projected) nearby star may require a correction of derived stellar and planetary parameters 
between a few to several tens percent \citep{Daemgen2009, Buchhave2011}.
  
Another important aspect of transiting planets is that they allow for a
measurement of the projected alignment between the orbital plane of the
planet and the rotational plane of the star through the
Rossiter-McLaughlin effect \citep[e.g. ][]{Queloz2000}. Studies using this
effect have shown that many transiting hot Jupiters are significantly
misaligned with respect to the stellar spin \citep[e.g. ][]{Winn2010},
which implies that three-body mechanisms such as Kozai migration
\citep[e.g. ][]{Fabrycky2007} may be responsible for the formation of those
systems, rather than classical orbital migration \citep[e.g. ][]{Lin1996}.
Since the Kozai mechanism requires the presence of a wide stellar
companion, it follows that searching for binarity among TEP hosts could
provide important clues for the formation of hot Jupiter systems.

In this paper we present high resolution Lucky Imaging observations of 21 TEP host stars using the two AstraLux instruments at the 2.2\,m telescope at
Calar Alto and at NTT at La Silla. The observations complement the sample of TEP hosts presented by \citet{Daemgen2009}, and are described in Sect. 2
together with the methods used for obtaining relative astrometry and photometry.
All the transiting exoplanets in our survey transit only one star, in a short-period orbit.
In Section 3 we derive magnitudes, $(i-z)$ colours, and photometric
spectral types and distances to the companion candidates from the photometric observations in 
SDSS $i^\prime$- and $z^\prime$-band and known spectral types of the planet host stars.
Evidence of physical companionship of the
planet host stars and the companion candidates is investigated, and the results for each individual target are compared with previously published 
astrometric and photometric data if available. We estimate the probability of chance alignment and perform a Bayesian analysis of our complete
observed sample of TEP hosts with AstraLux. The results are summarized and discussed in Sect. 4.

\section{Observations and data reduction}
\subsection{Observations with AstraLux}
The 21 transiting exoplanet host stars were observed within four different observing runs with the two AstraLux Lucky Imaging instruments. Most of the 
TEP hosts in the survey are located in the northern sky and were therefore observed with AstraLux Norte at the 2.2 m telescope
at Calar Alto observatory in October-November 2009, with follow-up astrometric observations in November 2011. 
The southern-sky targets were observed in November 2008 and April 2009
with the AstraLux Sur visitor instrument mounted to the ESO 3.5\,m New Technology Telescope (NTT) at La Silla \citep[see ][for details on the AstraLux instruments] 
{Hormuth2008,Hippler2009}. 
Lucky Imaging is a way to limit the effects of atmospheric turbulence by taking a large number of very short integrations from which only 
the least distorted few percent of the frames are selected. These are shifted and added to produce the final image, yielding almost diffraction-limited 
resolution.
The full AstraLux field of view (FoV) in the final resampled frames is $\approx15.7\arcsec\times15.7\arcsec$ for AstraLux Sur, and 
$\approx24\arcsec\times24\arcsec$ for AstraLux Norte.
The individual exposure time was either 15\,ms or 30\,ms, depending on the target brightness and observing
conditions. The shorter integrations were achieved by decreasing the FoV and reading out only 
a subframe of the detector. 
In order to match a total integration time of 300\,s, the number of integrations was set to 20\,000 or 10\,000 respectively.

Each target was observed in SDSS $i^{\prime}$- and $z^{\prime}$-filter. Astrometric reference stars in the open cluster NGC\,3603, the Orion 
Trapezium and/or the globular cluster 47 Tuc were 
observed each night for calibration of the detector rotation and pixel scale. 
Field rotation and plate scale were derived using custom IDL routines to compute separation and position angle of the astrometric reference stars
pairwise in several frames and compare to values derived from e.g. HST/WFPC2 archive images.
The plate scale for the observations with AstraLux Norte was 
$23.43\pm0.06$\,mas/px with the detector rotated $0.06\pm0.02\degr$ east of north in October-November 2009 and plate scale
$23.74\pm0.05$\,mas/px in November 2011 with a detector rotation of $1.6\pm0.2\degr$ to the east.
For the AstraLux Sur instrument we derived a plate scale of $15.373\pm0.002$\,mas/px and detector rotation $1.7\pm0.3\degr$ to the west of north
in November 2008 \citep[see ][]{Bergfors2010},
and $15.245\pm0.006$\,mas/px, $1.4\pm0.2\degr$ to the west in April 2009.

\begin{figure}
 \centering
   \includegraphics[width=8.6cm]{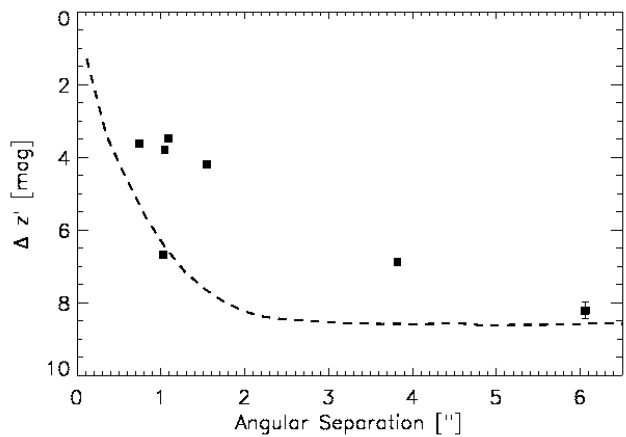}
    \caption{Average sensitivity of the AstraLux Norte observations in October-November 2009. The squares mark the companion candidates 
  (see Table \ref{binary}). 
The dashed line represents the typical 5$\sigma$ detection limit.}
\label{sensitivity}
 \end{figure}

\begin{figure*}
 \centering
   \includegraphics[width=18cm]{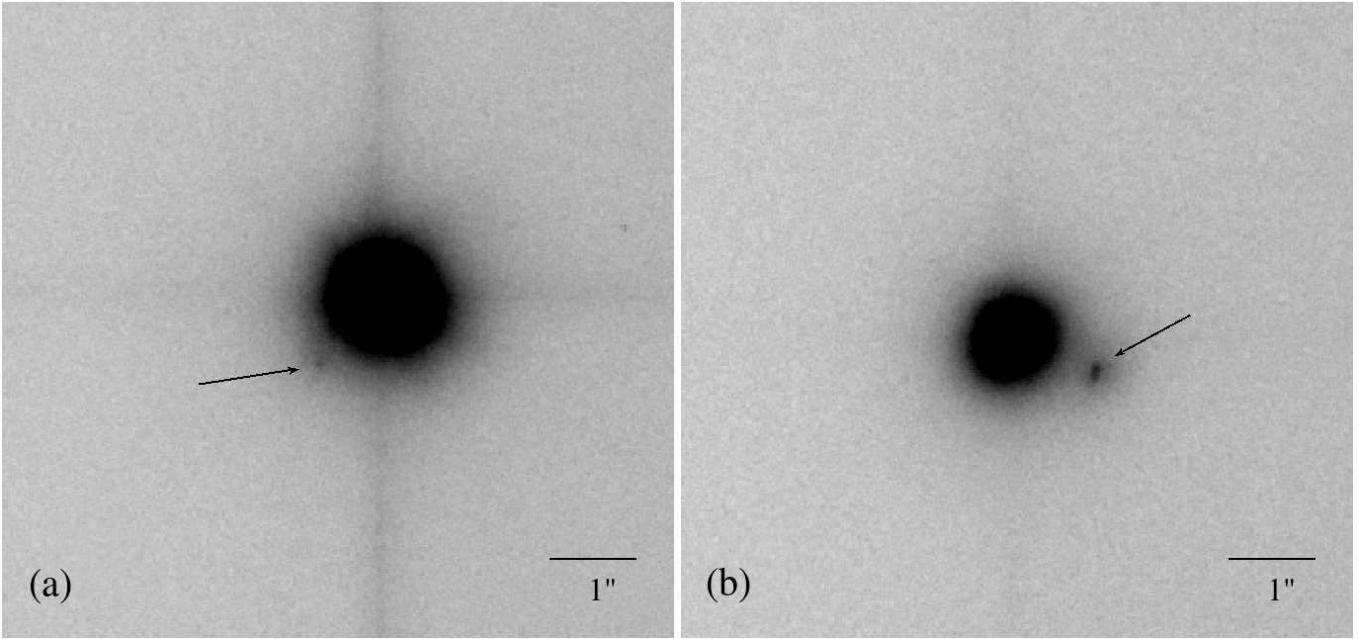}
    \caption{AstraLux Norte $z^{\prime}-$band observations of the candidate binaries (a) HAT-P-8 and (b) WASP-12 in October 2009. 
The images are shown in a square root scale with north up and east to the left. }
\label{HATP8_WASP12}
 \end{figure*}

\subsection{Photometry and astrometry}
In the sample of 21 TEPs, we found candidate companion stars from visual inspection of the reduced Lucky Imaging frames to 7 stars:
the previously known companion candidates to WASP-2 \citep{Cameron2007}, TrES-2 and TrES-4 \citep{Daemgen2009}, HAT-P-7 \citep{Narita2010}, and new
companion candidates to HAT-P-8, WASP-12 and XO-3. 
Stars for which no companions were detected within the FoV of AstraLux
in these observation runs are listed in Table \ref{single}. The stars for which we observed a candidate companion are listed in Table \ref{binary}.

Most of the companion candidates reside close to the primary stars, within the PSF-wings (HAT-P-7 and 
XO-3 being the exceptions). For these stars we performed relative photometry and astrometry of the companion candidates using mainly the IRAF \textit{allstar}
 \citep{Tody1986, Tody1993} task for PSF-fitting. The PSF was built from the primary star in an image where the secondary star had been removed
 \citep[see ][ for more details on the procedure]{Daemgen2009}.
Aperture photometry with IRAF \textit{apphot} was used to determine the properties of the wide companion candidates to XO-3 and HAT-P-7. 
The astrometric and photometric properties in Tables \ref{derived_properties},\ref{astrometry} are averaged measurements of the final images using the Lucky Imaging
combination of the best 5\% and 10\% of the individual integrations. For follow-up astrometry in November 2011, only $z'$-band measurements were used. 
The separation and position angle error bars are propagated from $1\sigma$ uncertainties in the positional measurements, and take into account 
systematic errors (i.e., plate scale and detector orientation, see Sect. 2.1). The uncertainty in detector rotation is usually the dominant
error source in the relative astrometry. The error bars in magnitude differences are propagated from the photometric 
errors estimated with IRAF. 

Figure \ref{sensitivity} 
shows the typical 5$\sigma$ detection limit for the observations with AstraLux Norte in October-November 
2009. At separation $\rho=0.5\arcsec$ we can typically detect a companion 4 magnitudes fainter than the primary in $\Delta z'$, and 6 magnitudes
fainter at $\rho=1\arcsec$. A similar
plot representing observations with AstraLux Sur can be found in \citet{Bergfors2010}. 
The $\Delta mag$ as a function of angular separation was estimated by pairwise subtraction of a set of images of single
stars with similar peak flux and FWHM from each other and measuring the noise level in concentric circles at increasing separations from 
the centre. 

\begin{table}
\caption{Transiting exoplanet host stars with no observed companions.}
\label{table:2}
\centering
\begin{tabular}{l l r r}
\hline\hline
Planet Host & Instrument & FoV & Obs. Date (UT) \\
	    &		& [\arcsec] & \\
\hline
HAT-P-2				&	AstraLux Norte	&	$12.00\times12.00$	&1 Nov 2009	\\
HAT-P-11			&	AstraLux Norte	&	$12.00\times12.00$	&29 Oct 2009	\\
HAT-P-13			&	AstraLux Norte	&	$12.00\times12.00$	&2 Nov 2009	\\
HD 149026			&	AstraLux Norte	&	$12.00\times12.00$	&1 Nov 2009	\\
HD 209458			&	AstraLux Sur	&	$7.87\times7.87$	&12 Nov 2008	\\
HD 80606\footnotemark[1]	&	AstraLux Norte	&	$12.00\times12.00$	&30 Oct 2009	\\
WASP-3				&	AstraLux Norte	&	$12.00\times12.00$	&30 Oct 2009	\\
WASP-4				&	AstraLux Sur	&	$15.74\times15.74$	&10 Nov 2008	\\
WASP-5				&	AstraLux Sur	&	$15.74\times15.74$	&10 Nov 2008	\\
WASP-7 				&	AstraLux Sur	&	$7.87\times7.87$	&10 Nov 2008	\\
WASP-13				&	AstraLux Norte	&	$12.00\times12.00$	&30 Oct 2009	\\
WASP-15				&	AstraLux Sur	&	$7.83\times7.83$	&12 Apr 2009	\\
XO-4				&	AstraLux Norte	&	$12.00\times12.00$	&1 Nov 2009	\\
XO-5				&	AstraLux Norte	&	$12.00\times12.00$	&31 Oct 2009	\\
\hline
\end{tabular}
\label{single}
\medskip
{$^{[1]}$ Visual binary star \citep[e.g., ][]{Naef2001}, outside of AstraLux field of view.}
\end{table}

\begin{table*}
\caption{Relative photometry and astrometry for the companion candidates.}
\label{table:3}
\centering
\begin{tabular}{l r l r r r r r}
\hline\hline
ID & Date of obs. & Instrument & FoV & Separation & Position Angle & $\Delta z'$ & $\Delta i'$ \\
   &    &            & [\arcsec] & [\arcsec]     & $[\degr]$       &  $[\rm mag]$    & $[\rm mag]$   \\     
\hline
HAT-P-7\footnotemark[1]	&	30 Oct 2009	&	AstraLux Norte	&	12.00$\times$12.00	&	3.82$\pm$0.01	&	90.4$\pm$0.1	&	6.89$\pm$0.07	&	7.94$\pm$0.21	\\
			&	9 Nov 2011	&	AstraLux Norte	&	12.00$\times$12.00	&	3.86$\pm$0.07	&	89.9$\pm$0.8	&	6.87$\pm$0.07	&	7.58$\pm$0.07	\\
HAT-P-8			&	29 Oct 2009	&	AstraLux Norte 	&	12.00$\times$12.00	&	1.027$\pm$0.011	&	137.3$\pm$0.4	&	6.68$\pm$0.07	&	7.34$\pm$0.10	\\
TrES-2\footnotemark[2]	&	29 Oct 2009	&	AstraLux Norte 	&	12.00$\times$12.00	&	1.085$\pm$0.006	&	136.1$\pm$0.2	&	3.48$\pm$0.06	&	3.73$\pm$0.03	\\
TrES-4\footnotemark[2]	&	30 Oct 2009	&	AstraLux Norte 	&	12.00$\times$12.00	&	1.550$\pm$0.007	&	359.9$\pm$0.2	&	4.19$\pm$0.05	&	4.57$\pm$0.05	\\
WASP-2\footnotemark[3]	&	13 Apr 2009	&	AstraLux Sur	&	15.66$\times$15.66	&	0.761$\pm$0.009	&	103.5$\pm$0.2	&	3.62$\pm$0.05	&	… 		\\
			&	29 Oct 2009	&	AstraLux Norte	&	12.00$\times$12.00	&	0.739$\pm$0.024	&	104.0$\pm$1.3	&	3.64$\pm$0.04	&	4.17$\pm$0.03	\\
			&	9 Nov 2011	&	AstraLux Norte	&	12.00$\times$12.00	&	0.744$\pm$0.013	&	104.6$\pm$0.7	&	…		&	…		\\
WASP-12			&	30 Oct 2009	&	AstraLux Norte	&	12.00$\times$12.00	&	1.047$\pm$0.021	&	249.7$\pm$0.8	&	3.79$\pm$0.10	&	4.03$\pm$0.07	\\
			&	8,9 Nov 2011	&	AstraLux Norte	&	12.00$\times$12.00	&	1.043$\pm$0.014	&	249.9$\pm$0.5	&	…		&	…		\\
XO-3			&	30 Oct 2009	&	AstraLux Norte	&	12.00$\times$12.01	&	6.059$\pm$0.047	&	296.7$\pm$0.3	&	8.22$\pm$0.23	&	8.57$\pm$0.24	\\
\hline
\end{tabular}
\label{binary}
\medskip

{$^{[1]}$ The October 2009 observations were first published in \citet{Narita2010}}
{$^{[2]}$ Companion was discovered by \citet{Daemgen2009}}
{$^{[3]}$ Companion was discovered by \citet{Cameron2007} and has also been observed by \citet{Daemgen2009}}
\end{table*}

\begin{table*}
\caption{Photometric properties of the primary stars and the companion candidates.}
\centering
\begin{tabular}{l r r r r r r r r r}
\hline\hline
Planet Host & SpT(A) & SpT(B) & $\rm m_{z}(A)$ & $\rm m_{i}(A)$ & $\rm m_{z}(B)$ & $\rm m_{i}(B)$ & $\rm (i-z)_B$ & d(A) & d(B) \\
	    &        &        &   [mag]       &     [mag]     &     [mag]     &  [mag]        & 	[mag]    & [pc] & [pc]   \\     
\hline
HAT-P-8\footnotemark[1]	& F8\,V	& M1\,V ... M3.5\,V 	& 10.20$\pm$0.1	& 10.18$\pm$0.1	& 16.88$\pm0.1$	& 17.52$\pm0.1$	&0.64$\pm0.14$	& 150$\pm$20	& 710 ... 350	\\
			& F5\,V	& M1\,V ... M3.5\,V	& 10.08$\pm$0.1	& 10.03$\pm$0.1	& 16.76$\pm0.1$	& 17.37$\pm0.1$	&0.61$\pm0.14$	& 190$\pm$20	& 670 ... 330	\\
TrES-2			& G0\,V	& K0\,V ... M0\,V	& 11.11$\pm$0.1	& 11.11$\pm$0.1	& 14.59$\pm0.1$	& 14.84$\pm0.1$	&0.25$\pm0.14$	& 220$\pm$20	& 670 ... 270	\\
TrES-4 			& F8\,V	& K4.5\,V ... M1.5\,V	& 11.57$\pm$0.1	& 11.55$\pm$0.1	& 15.76$\pm0.1$	& 16.12$\pm0.1$	&0.36$\pm0.14$	& 290$\pm$30	& 790 ... 380	\\
WASP-2\footnotemark[2]	& K0\,V	& M1\,V ... M3.5\,V	& 11.10$\pm$0.1	& 11.20$\pm$0.1	& 14.74$\pm0.1$	& 15.37$\pm0.1$	&0.63$\pm0.14$	& 140$\pm$10	& 260 ... 130	\\
			& K2\,V	& M1.5\,V ... M4\,V	& 11.19$\pm$0.1	& 11.33$\pm$0.1	& 14.83$\pm0.1$	& 15.50$\pm0.1$	&0.67$\pm0.14$	& 120$\pm$10	& 240 ... 110	\\
WASP-12			& G0\,V	& K0\,V ... M0\,V	& 11.41$\pm$0.1	& 11.41$\pm$0.1	& 15.20$\pm0.1$	& 15.44$\pm0.1$	&0.24$\pm0.14$	& 250$\pm$30	& 890 ... 350	\\
XO-3\footnotemark[3]	& F5\,V	& G0\,V ... M2.5\,V	& 9.91$\pm$0.1	& 9.86$\pm$0.1	& 18.13$\pm0.2$	& 18.43$\pm0.2$	&0.30$\pm0.30$	& 170$\pm$20	& 5470 ...860 	\\

\hline
\label{derived_properties}
\end{tabular}
\medskip
{$^{[1]}$ A primary star spectral type of F8\,V is suggested by \citet{JonesSleep}, while we find F5\,V an equally good fit to the SEDs
of \citet{KrausHillenbrand2007}. We therefore list both alternatives here.}
{$^{[2]}$ The primary star is of spectral type K1\,V. \citet{KrausHillenbrand2007} provides spectral type models only for K0\,V and K2\,V, and we choose to
present these two alternatives rather than interpolate between spectral types.}
{$^{[3]}$ Here we assume that the companion candidate is a main sequence star when deriving spectral type and distance, however see Sect. 3.2.6}
\end{table*}

\begin{table}
\caption{Summary of astrometric measurements.}
\label{astrometry}
\centering
\begin{tabular}{l l l l r}
\hline\hline
Planet Host & Date of Obs. & Separation & Pos. Ang. & Ref.\\
            &              & [\arcsec]         & [\degr]            &  \\
\hline
HAT-P-7	&	6 Aug 2009	&	3.88$\pm$0.01	&	89.8$\pm$0.3	&	[1]	\\
	&	30 Oct 2009	&	3.82$\pm$0.01	&	90.4$\pm$0.1	&	[1],[2]	\\
	&	9 Nov 2011	&	3.86$\pm$0.07	&	89.9$\pm$0.8	&	[2]	\\
HAT-P-8	&	29 Oct 2009	&	1.027$\pm$0.011	&	137.3$\pm$0.4	&	[2]	\\
TrES-2	&	May 2007	&	1.089$\pm$0.008	&	135.5$\pm$0.1	&	[3]	\\
	&	29 Oct 2009	&	1.085$\pm$0.006	&	136.1$\pm$0.2	&	[2]	\\
TrES-4	&	Jun 2008	&	1.555$\pm$0.005	&	359.8$\pm$0.1	&	[3]	\\
	&	30 Oct 2009	&	1.550$\pm$0.007	&	359.9$\pm$0.2	&	[2]	\\
WASP-2	&	Nov 2007	&	0.757$\pm$0.001	&	104.7$\pm$0.3	&	[3]	\\
	&	13 Apr 2009	&	0.761$\pm$0.009	&	103.5$\pm$0.2	&	[2]	\\
	&	29 Oct 2009	&	0.739$\pm$0.024	&	104.0$\pm$1.3	&	[2]	\\
	&	9 Nov 2011	&	0.744$\pm$0.013	&	104.6$\pm$0.7	&	[2]	\\
WASP-12	&	30 Oct 2009	&	1.047$\pm$0.021	&	249.7$\pm$0.8	&	[2]	\\
	&	8,9 Nov 2011	&	1.043$\pm$0.014	&	249.9$\pm$0.5	&	[2]	\\
\hline
\end{tabular}
\medskip
{$^{[1]}$ \citet{Narita2010}}
{$^{[2]}$ This paper}
{$^{[3]}$ \citet{Daemgen2009}}
\end{table}

\section{Results}
\subsection{Properties of the stellar companion candidates}
The spectral types of the companion candidates are derived using primary star spectral types from literature and assuming that the
additional flux from the much fainter secondary stars do not affect these spectral types. We use stellar SEDs by \citet{KrausHillenbrand2007}
for the known TEP host spectral type in combination with 2MASS photometry in the JHK bands \citep{Cutri2003} to find the average distance modulus and
SDSS $i$- and $z$-band magnitudes.
The measured magnitude difference between the primary and secondary stars in $i^{\prime}$
and $z^{\prime}$ then provides apparent magnitudes and ($i-z$) colour for the companion candidate. 
While we do not explicitly correct for the 
transformation between SDSS $i^{\prime},z^{\prime}$ to $i,z$, 
the difference, when calculated from the photometric transformation equations (SDSS webpage),
is small. The magnitude difference between the two photometric systems is less than $(i-i^{\prime})=0.05$ mag for stars bluer than $(r-i)\approx1.5$, 
and even less in $(z-z^{\prime})$. According to the SEDs of \citet{KrausHillenbrand2007}, this $(r-i)$ colour corresponds to spectral types 
M4 or earlier. None of our derived 
spectral types are later than M4, and we conservatively assume error bars of $\pm0.1$ for the derived $i,z$ magnitudes of the primary and secondary stars
so as to include the transformation between photometric systems as well as uncertainty in the photometric measurements ($\pm0.2$ for the
very faint companion candidate to XO-3). 
The secondary star spectral types and primary and secondary photometric distances were estimated from the derived $(i-z)$ colour and the 
SEDs of \citet{KrausHillenbrand2007}. Interstellar extinction and stellar metallicity were not considered in these
estimates but may affect the photometric distances, as well as the colours and spectral types. 
The photometric distance estimates are therefore only indicative and not hard limits.
The components' spectral types, apparent magnitudes, 
photometric distances and the secondary stars' $(i-z)$ colours are listed in Table \ref{derived_properties}.

\subsection{Notes on individual systems}
Companion candidates to 7 transiting exoplanet hosts were found from these observations. Of these, the three TEP hosts 
WASP-2, TrES-2 and TrES-4 and their candidate companions had been observed with AstraLux previously \citep{Daemgen2009}, and the candidate 
companion to HAT-P-7 is discussed in \citet{Narita2010}.
The widely separated candidate companion to XO-3 is likely to be a physically unrelated background object if it is a main sequence star, 
although the possibility of a coeval white dwarf companion can not be ruled out from these observations.
The two faint objects at separations of $\rho\sim1\arcsec$ 
to HAT-P-8 and WASP-12 are previously unknown, plausibly bound stellar companions. Figure
\ref{HATP8_WASP12} shows the AstraLux Norte $z^{\prime}-$band observations of these two systems obtained in October 2009. All astrometric
measurements including previous observations are summarised in Table \ref{astrometry}.

\subsubsection{HAT-P-8}
The transiting exoplanet HAT-P-8 b was discovered by \citet{Latham2009}. It is a slightly inflated planet with mass $\rm M_p=1.52^{+0.18}_{-0.16}M_J$ and
radius $\rm R_p=1.50^{+0.08}_{-0.06}R_J$ \citep{Latham2009}.
The host star spectral type is only given as F in the discovery paper, but is referred to as F8\,V by \citet{JonesSleep}. We find from the 2MASS
JHK photometry and colours that a spectral type of F5\,V also fits the SEDs of \citet{KrausHillenbrand2007}, and list both alternatives in 
Table \ref{derived_properties} until a more precise spectral classification can be made. Although the assumed primary spectral type affects the 
derived primary $i',z'$ magnitudes and thereby the colour and spectral type of the companion candidate (see Sect. 3.1,) we find for both alternatives 
of primary spectral type that the stellar candidate companion is likely to be of spectral type M2 to M4 from the $(i-z)$-colours. 

\subsubsection{TrES-2}
The companion candidate was first discovered by \citet{Daemgen2009} from AstraLux Norte observations in May 2007. The observations presented here
took place in October 2009. We measure a separation of $\rm \Delta RA=0.752\arcsec\pm0.007\arcsec, \Delta Dec=0.782\arcsec\pm0.007\arcsec$, which is 
consistent with the astrometry of \citet{Daemgen2009} who found $\rm \Delta RA=0.763\arcsec\pm0.007\arcsec, \Delta Dec=0.777\arcsec\pm0.007$.
The proper motion of TrES-2 is only $\rm \mu_\alpha\cos\delta=2.34\pm1.7\,mas/yr, \mu_\delta=-1.55\pm1.7\,mas/yr$ \citep[PPMX Catalog, ][]{Roeser2008}, 
which is within our positional error bars over the time interval of $\sim2.4$\,years between observations. We have to await future observations to tell 
whether the pair is physically bound or not. 

\subsubsection{TrES-4}
The candidate companion star was discovered in AstraLux Norte observations from June 2008 by \citet{Daemgen2009}. Our measurement of 
$\rm \Delta RA=0.003\arcsec\pm0.006\arcsec, \Delta Dec=1.550\arcsec\pm0.007\arcsec$
in October 2009 is consistent with the separation 
$\rm \Delta RA=0.005\arcsec\pm0.003\arcsec, \Delta Dec=1.555\arcsec\pm0.005\arcsec$
measured by \citet{Daemgen2009}.
With a proper motion of $\rm \mu_\alpha\cos\delta=-9.94\pm2.5\,mas/yr, \mu_\delta=-27.80\pm2.5\,mas/yr$ \citep{Roeser2008},
the observations over a time baseline of 16 months differ by $\approx1.4\sigma$ in RA and $\approx4.3\sigma$ in Dec from what would be
expected if the companion candidate was 
a physically unrelated stationary background star, where $\sigma$ is the quadrature sum of the observed and proper motion errors. 
We thus conclude that this is a common proper motion binary.

\subsubsection{WASP-2}
WASP-2\,b was discovered by \citet{Cameron2007}, who also reported the stellar companion candidate to the east of the TEP host star 
at an angular separation of $\rho=0.7\arcsec$.
The first AstraLux observation of this target was obtained in November 2007 \citep{Daemgen2009}.
Although the primary star is a K1\,V star, we list in Table \ref{derived_properties} the derived magnitudes, colours and photometric distances 
assuming primary spectral type K0 and K2, since colours for
K1 are not provided by \citet{KrausHillenbrand2007}. Depending on primary spectral type (see Sect. 3.1), we find that the companion candidate is of 
spectral type M1/M2 ... M4. \citet{Daemgen2009} observed the companion candidate at separation 
$\rm \Delta RA=0.732\arcsec\pm0.004\arcsec, \Delta Dec=0.192\arcsec\pm0.004\arcsec$ in November 2007. In November 2011 we observed the separation
$\rm \Delta RA=0.720\arcsec\pm0.015\arcsec, \Delta Dec=0.188\arcsec\pm0.012\arcsec$.
The proper motion of the planet host star is $\rm \mu_\alpha\cos\delta=3.38\pm2.9\,mas/yr, \mu_\delta=-52.31\pm2.9\,mas/yr$. The observations over
a time baseline of 4 years are marginally inconsistent with the hypothesis of a stationary background object in RA ($\approx1.3\sigma$) but by more
than $11\sigma$ in Dec. We therefore conclude that this is a common proper motion system.

\subsubsection{WASP-12} The very bloated planet WASP-12\,b was discovered by \citet{Hebb2009}. It is a highly irradiated planet, and one of the hottest 
with an equilibrium temperature $\rm T_{eq}=2516\pm36$\,K \citep{Hebb2009}.

The stellar companion candidate is not previously known. We derive a spectral type K2 ... M0\,V from the measured ($i-z$) colour and the 
SEDs of \citet{KrausHillenbrand2007}. 
The proper motion of WASP-12 is $\rm \mu_\alpha\cos\delta=-0.36\pm1.7\,mas/yr, 
\mu_\delta=-6.38\pm1.7\,mas/yr$ \citep{Roeser2008}, and future observations will be necessary to determine whether or not the stars are physical 
companions.
The companion candidate has an elongated shape (Fig. \ref{HATP8_WASP12}) on both dates of observation, which suggests that the companion
candidate may be an unresolved binary.

\subsubsection{XO-3}
A very faint companion candidate ($\Delta z\approx8.2$) was found at large angular separation ($\rho\approx6\arcsec$) from the TEP host star XO-3. 
The $(i-z)$-colour places the companion candidate at a distance of $\ga860$\,pc if a main sequence star, 
and it is hence likely to be a non-related background object. 
Another possibility is that the companion candidate is a white dwarf at the approximately same distance as the TEP host. The colours and
brightness are consistent with a white dwarf with $\rm T_{eff}\approx4000$\,K, corresponding to a very hydrogen deficient white dwarf with 
cooling age 4\,Gyr according to cooling curves by \citet{ChenHansen2011}. 
The age of XO-3 has been estimated to $2.82_{-0.82}^{+0.58}$\,Gyr \citep{Winn2008}, and is thus compatible with a 4.6\,Gyr white dwarf at 
$3\sigma$. 
While the candidate is unlikely to be a coeval comoving white dwarf companion, we can not exclude the possibility until future proper motion 
observations can be performed.

\subsection{Probability of chance alignment}
The probability of chance alignment is estimated using the statistical approach of \citet{Daemgen2009}.
The density of detectable background giants, $\rho\left(m_K\right)$, is calculated by selecting all stars included in the 2MASS PSC \citep{Cutri2003} 
within 30\arcmin of each of the observed targets (Tables \ref{single}, \ref{binary}) that are brighter than the estimated limiting magnitude of AstraLux,
$m_K\approx14$, and redder than $\left(J-K\right)\ge0.5$. The probability of detecting a background giant is

\begin{equation}\label{prob}
   P\left(\Theta,m_K\right)=1-e^{-\pi\rho\left(m_K\right)\Theta^2},
\end{equation} 

where $\Theta$ is the maximum angular separation \citep{Brandner2000}. Using the aforementioned cuts in $m_K$, $\left(J-K\right)$ for $\Theta=2\arcsec$
\citep[see ][]{Daemgen2009} the average probability of finding a non-related background star within 2\arcsec to the target star is $P=0.08\%$.
We would then expect to detect unrelated background sources to $E=21\times P\approx0.016$ of our observed targets.
This expectation value of chance alignment increases to 0.14 at the separation of 6\arcsec at which we find the probable background object
in the observations of XO-3 (see Sect. 3.2.6). 

We also consider a possible contamination by M dwarfs by selecting stars in the 2MASS PSC with $m_K\le14.0$\,mag, and NIR colours J-H 
$\ge$0.55\,mag and H-K $\ge$0.0\,mag. We get a 
typical surface density of 6$\times$10$^{-5}$ M dwarfs per square arcsec, which corresponds to a probability of 7.5$\times$10$^{-4}$ to find an 
unrelated M dwarf within $2\arcsec$ of one of our target stars. We conclude that the probability of contamination by an unrelated M dwarf is of the same
order as contamination by a background giant.
The close candidate companions in our sample
are thus likely to be true companions, although future observations are still necessary for confirmation of common proper motions for HAT-P-8,
TrES-2, HAT-P-7 and WASP-12.

\subsection{Bayesian statistical analysis}

We adapt the Bayesian statistical analysis method established in \citet{Allen2007} and utilized as well in \citet{Kraus2011} to our survey for binarity
of exoplanet host stars. We model the distributions of stellar binary mass ratios and projected separations as a power law in mass ratio 
($\rm q=M_{sec}/M_{prim}$), and a Gaussian
in projected separation.
Thus, we are left with 4 parameters to our models: 

\begin{itemize}
\item F -- the companion star fraction (number of companions per primary star)
\item $\gamma$ -- the power law index for the mass ratio (q) power law
\item r$_{0}$ -- the average radius, i.e. the center of the Gaussian
  of projected separations
\item $\sigma$ -- the width of the Gaussian of projected separations
\end{itemize}

The full combined form of our model is:

\begin{equation}
R(q, r | F, \gamma, r_{o}, \sigma ) \propto F q^{\gamma} exp(- \frac{(r - r_{o})^{2}}{2\sigma^{2}} )
\end{equation}

This equation gives us the probability of detecting a binary companion to a given star in a given projected separation, mass ratio range given a 
particular set of values for $F$, $\gamma$,
 $r_0$ and $\sigma$. 

To calculate the likelihood, we must compare this model value with the actual data in each part of parameter space. To do so, we must calculate 
how many binary companions we expect to detect with this model in each projected separation, mass ratio bin and then compare with the actual number 
detected in each bin (generally 0!). For a given mass ratio, projected separation bin and set of model
parameters, the number of companions predicted will be:

\begin{equation}
N_{pred} (q,r) = N_{obs} R(q,r | F, \gamma, r_{o}, \sigma )
\end{equation}  

where N$_{obs}$ is the number of times this projected separation, mass ratio bin was observed in our survey (derived from the contrast curves and
stellar properties of each survey star). To compare data and model, we need to adopt a likelihood estimator.
Since we expect to detect only small numbers of binary companions, our survey can be treated as a counting experiment. Thus, we adopt
Poisson statistics here to calculate the likelihood:

\begin{equation}
likelihood = prob(N_{det} | F, \gamma, r_{o}, \sigma ) = \frac{N_{pred}^{N_{det}} exp(-N_{det})}{N_{det}!}
\end{equation}

To derive the posterior probablitity distribution function (PDF) for this bin, we must multiply the likelihood by any prior probability distribution for
 our parameters. Certain priors can be adopted according to the functional form of the model parameters.
We adopt the simplest uniform priors for $\gamma$ and r$_{o}$:

\begin{equation}
prob(\gamma|I) = prob( r_{o} | I) = 1
\end{equation}

However, F and $\sigma$ are scale parameters (i.e. invariant to changes in scale), allowing us to adopt a slightly more complex prior:

\begin{equation}
prob(F|I) = \frac{1}{F}
\end{equation}

\begin{equation}
prob( \sigma | I) = \frac{1}{\sigma}
\end{equation}

Thus our full prior is:

\begin{equation}
prior = prob( F, \gamma, r_{o}, \sigma | I ) = \frac{1}{F \sigma}
\end{equation}

For a justification of this choice, please see pages 109-110 in \citet{SiviaSkilling2006}.  Multiplying the likelihood and 
prior then yields the posterior PDF for this projected separation, mass ratio bin.
This can be generalized across all projected separation, mass ratio bins
for the survey fairly easily.  We generalize N$_{obs}$ and N$_{det}$ into 2d arrays for each separation, mass ratio
bin observed, which we will henceforth call the window function and detection array.  

To build the window array, we use the contrast curve for each survey star to define the ranges in angular separation and contrasts where binary
companions can be detected and the ranges where the contrast is insufficient to do so.
Bins where a companion can be detected are assigned a value of 1; bins where no companion can be detected are assigned a value of 0.
The window function for each star is then converted to projected separation vs. minimum detectable mass ratio
using the models of \citet{Baraffe1998} and the distance to each survey star.  z'-band contrasts and magnitudes
were converted to H and K band for comparison with the \citet{Baraffe1998} models by interpolating the stellar SEDs 
from Table 5 from \citet{KrausHillenbrand2007}. 
All the single star window functions are added together to form the survey window function. The detection array is set up in a similar manner --- first 
as a simple array with the number of objects detected in each separation
and mass bin, and then deprojected into projected separation, mass ratio space.

\begin{figure}
 \centering
   \includegraphics[width=8.6cm]{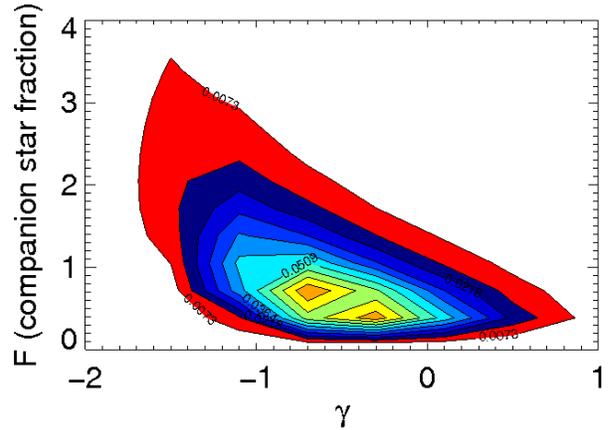}
    \caption{2-D marginalized PDF for mass ratio power law index $\gamma$ vs. companion star fraction F. }
\label{gamma-F}
 \end{figure}

\begin{figure}
 \centering
   \includegraphics[width=8.6cm]{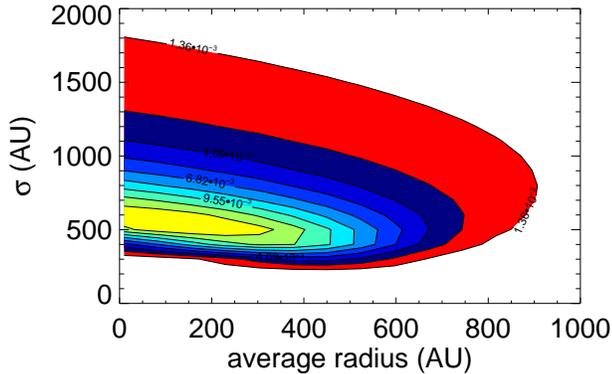}
    \caption{2-D marginalized PDF for the average projected binary separation $r_0$ vs. the width of the Gaussian of projected separations $\sigma$. }
\label{rad-sigma}
 \end{figure}

Then, for each set of model parameter, we calculate for each bin the posterior PDF. We multiply the posterior PDFs from each bin of
observable space together to get the full posterior PDF across observable space for this set of model parameters.  This process is repeated for all 
sets of model parameters of interest to derive the full posterior PDF as a function of the four model parameters.  
We calculated the posterior PDF for a grid in projected separation and mass ratio, running from 10 AU to 2000 AU in steps of 10 AU for projected
separation and from 0.01 to 1.0 in q, in steps of 0.005.  We allowed $\gamma$ to run from -3.9 to 3.7, in steps of 0.4 and F to run from 0.05 to 9.72 
in steps of 0.33.  The average radius, r$_{o}$, ran from 10 AU to 2137 AU in steps of 73.33 AU and $\sigma$ ran from 1 to 2901 AU in steps of 100 AU.  
Because it is not possible to visualize the full 4-dimensional posterior PDFs, we plot 2-d marginalized posterior PDFs in the $\gamma$--F and 
$r_0$--$\sigma$ plane in Figures \ref{gamma-F},\ref{rad-sigma}. 
The maximum of the posterior PDF yields the most-likely combination of model parameters given the information in hand.
In this case, the maximum posterior value occurs with the parameter combination: r$_{o}$ = 377 AU, $\sigma$=401 AU, F = 0.38, and
$\gamma$=-0.7.  However, from the marginalized posterior PDFs shown in Figures \ref{gamma-F}, \ref{rad-sigma}, it is clear that many combinations of 
model parameters have fairly high likelihoods and that the covariance between parameters is fairly large.  At the 67$\%$ (90$\%$) confidence 
level, we constrain 
r$_{o}$ to lie between 10\,AU and 890\,AU (10 -- 1476\,AU), $\sigma$ to lie between 300 and 1800\,AU (300 -- 2901\,AU), 
$\gamma$ to lie between -1.1 and 0.5 (-1.5 -- 0.9) and 
F to lie between 0.38 and 2.05 (0.38 -- 4.72).

\section{Discussion}
The presence of a close binary companion is thought to affect processes of formation and subsequent dynamical evolution of 
the planets in such a system. We performed high angular resolution Lucky Imaging of 21 host stars to transiting exoplanets in search of faint, close
stellar companions. Two previously unknown companion candidates were discovered, to the TEP hosts HAT-P-8 and WASP-12. Future follow-up observations are
necessary to confirm common proper motion and hence physical companionship. Of the three candidate binary TEP hosts observed previously with AstraLux
\citep{Daemgen2009},
TrES-4 and WASP-2 were confirmed as common proper motion couples from follow-up observations presented in this paper. Observations over a longer 
time baseline are still necessary to confirm the TrES-2 system as a physical binary, although we find that the probability of non-related background
objects is very low (see Sect. 3.3). A wide companion candidate to XO-3 is likely to be a background object, although we cannot rule out the
possibility of a comoving white dwarf companion from these observations.

We performed a Bayesian analysis of the full sample of 31 TEP hosts observed with AstraLux, including the sample presented in \citet{Daemgen2009}.
The companion star fraction of the TEP host stars is not significantly different from that of solar type field stars 
\citep{DuquennoyMayor1991, Raghavan2010}. 
This is in accordance with what was found in previous compilations of multiplicity among RV planet hosts 
\citep{Raghavan2006, DesideraBarbieri2007, BonavitaDesidera2007}. At the 67\% confidence level we find a companion star fraction lower limit of 
$F\ge0.38$ which is slightly higher than the lower limits found by \citet{Raghavan2006} of $\approx26\%$ and \citet{Mugrauer2009}, $\approx20\%$.
However, 
those multiplicity fractions are based on samples of RV planet hosts and do not account for the common RV target selection bias of excluding known
binaries from exoplanet surveys. No such pre-selection of single targets is made in transit surveys, and may explain the higher fraction derived
in this survey. 

The slope of the mass-ratio distribution is likely to be uniform or slightly negative with a maximum posterior PDF value of $\gamma=-0.7$ (67$\%$
 confidence), 
indicating a preference for low mass ratio and hence large magnitude difference between companion stars.
While this result is again in agreement with the field population of solar-type stars \citep{DuquennoyMayor1991,Raghavan2010}, the blending by a nearby, 
bright star may cause rejection in transit candidate follow ups, and thus the sample of TEP hosts may be biased towards faint, low mass-ratio 
companions. 

The main difference in multiplicity properties between field stars and planet host stars from this survey appears to be the average binary
separation, which is about a factor of ten larger for the TEP hosts. With AstraLux we are typically sensitive to companions 2 magnitudes fainter than 
the planet host at angular separation of $\rho=0.2\arcsec$, corresponding to 40\,AU at the average TEP host distance of 200\,pc in our target sample.
All our discovered candidate companions are found at projected separations $\ge100$\,AU. While our sample is still small, the lack of close, faint
companions may be a real property of exoplanet hosts.

\section*{Acknowledgments}
The authors thank the reviewer, Philippe Delorme, for very useful comments and suggestions that improved the clarity of this paper.
We thank the staff at the Calar Alto and La Silla observatories for their support. M. J. acknowledges support by the Hubble fellowship.
This publication has made use of the SIMBAD database, operated at CDS, Strasbourg, France, and the Extrasolar Planets Encyclopaedia
maintained by Jean Schneider (http://exoplanet.eu).

\bibliographystyle{mn2e}
\bibliography{Bergfors_TEPs}

\label{lastpage}

\end{document}